\documentclass[12pt]{article}
\usepackage{amsmath,amssymb,bbold,epsfig}
\usepackage{amsfonts}
\usepackage{slashed}
\usepackage{dsfont}
\usepackage{cite}
\usepackage{graphicx,color}
\usepackage{bbm}
\usepackage[usenames,dvipsnames]{xcolor}
\definecolor{lightgray}{gray}{0.90}
\usepackage[colorlinks=true,linkcolor=red,citecolor=green,urlcolor=cyan,
bookmarks=true,bookmarksopen=true,pdfpagemode=None,pdfstartview=FitH]{hyperref}
\newcommand{\appsection}{\addtocounter{section}{1}\setcounter{equation}{0}
                         \renewcommand{\thesection}{\Alph{section}}
}
\renewcommand{\theequation}{\arabic{equation}}
\newcommand{\be}{\begin{equation}}
\newcommand{\ee}{\end{equation}}
\newcommand{\bea}{\begin{eqnarray}}
\newcommand{\eea}{\end{eqnarray}}

\begin{document}

\title{
\vglue 1.3cm
\Large \bf
Do non-relativistic neutrinos 
oscillate?}
\author{
\vspace*{-0.2cm}
{Evgeny~Akhmedov\footnote{Also at the National Research Centre 
Kurchatov Institute, Moscow, Russia}~\thanks{Email: \tt 
akhmedov@mpi-hd.mpg.de}
\vspace*{5.5mm}
} \\
{\normalsize\em 
Max-Planck-Institut f\"ur Kernphysik, Saupfercheckweg 1, 
}\\
{\normalsize\em
69117 Heidelberg, Germany
\vspace*{0.85cm}} 
}
\date{}  

\maketitle 
\thispagestyle{empty} 
\vspace{-0.8cm} 
\begin{abstract} 
We study the question of whether oscillations between non-relativistic 
neutrinos or between relativistic and non-relativistic neutrinos are possible. 
The issues of neutrino production and propagation coherence and their impact 
on the above question are discussed in detail. 
It is demonstrated that no neutrino oscillations can occur when 
neutrinos that are non-relativistic in the laboratory frame are involved, 
except in a strongly mass-degenerate case. We also discuss how this analysis 
depends on the choice of the Lorentz frame. Our results are for the most part 
in agreement with Hinchliffe's rule. 
\end{abstract}
\vspace{0.2cm}
\centerline{Pacs numbers: 14.60.Pq, 03.65.-w}
\vspace{0.3cm}

\newpage


\section{\label{sec:intro}Introduction} 
In virtually all theoretical studies of neutrino oscillations it is assumed 
that neutrinos are extremely relativistic, and for good reason: we 
normally do not deal with non-relativistic neutrinos in our everyday life. 
This does not mean, however, that non-relativistic neutrinos do not exist in 
nature. From neutrino oscillation experiments it follows that 
either two mass-eigenstate neutrino species have mass $\gtrsim 0.05$ eV, or 
one of them has mass $\gtrsim 0.05$ eV and one other 
$\gtrsim 8\times 10^{-3}$ eV. The present-day temperature of the cosmic 
neutrino background $T_\nu\simeq 1.95$\,K $\simeq 1.68\times 10^{-4}$ eV 
therefore means that at least two relic neutrino species are currently 
non-relativistic. Moreover, there are some indications of possible existence 
of (predominantly) sterile neutrinos with an eV-scale mass, and 
chiefly sterile neutrinos with keV -- MeV -- GeV scale masses 
are also being discussed \cite{white1,drewes,white2}. If exist, such neutrinos 
may well be non-relativistic in some situations of practical interest. The 
question of whether non-relativistic neutrinos can oscillate is therefore not 
of completely academic nature. In addition, as we shall see, 
it is related to some fundamental aspects of the theory of neutrino 
oscillations and therefore is of considerable conceptual interest.

When asking whether non-relativistic neutrinos can oscillate, 
one should obviously specify the reference frame in which neutrinos are 
considered. Neutrinos that are non-relativistic in one Lorentz frame may be 
highly relativistic in another, and vice versa. If not otherwise specified, 
we shall be considering neutrinos in the laboratory frame, by which we mean 
the frame in which the neutrino source is at rest or is slowly moving.%
\footnote{This definition may sometimes differ from what one would normally 
consider to be the laboratory frame in a neutrino experiment. For instance, in 
accelerator neutrino experiments the conventional definition would likely be 
that this is the frame in which the neutrino detector is at 
rest, whereas neutrinos are produced in decays of relativistic pions. 
} 
For neutrinos produced in decays, the source is just the parent particle; for 
neutrinos produced in collisions, the definition of the velocity of the source 
is more complicated and involves a consideration of the velocities and 
wave packet lengths of all particles participating in neutrino production 
\cite{beuthe}. 

The question of whether non-relativistic neutrinos can take part in neutrino 
oscillations has been discussed in 
refs.~\cite{nonrel1,nonrel2,nonrel3,nonrel4,nonrel5,beuthe, 
nonrel6}, with varying degree of detail and differing 
conclusions. In refs.~\cite{nonrel1,nonrel2,beuthe} it was argued that 
non-relativistic neutrinos cannot 
participate in neutrino oscillations, whereas 
in refs.~\cite{nonrel3,nonrel4,nonrel5,nonrel6} the opposite conclusion 
has been reached (or the opposite was implicitly assumed). 
In this paper we demonstrate that, unless non-relativistic neutrinos are 
highly degenerate in mass, their large energy and momentum differences 
prevent coherence of different neutrino mass eigenstates and therefore 
preclude flavour oscillations. 
Some of the arguments presented here have already been discussed in the 
literature, though mostly at a qualitative level. We put them on a 
more quantitative basis in this paper. In addition, we discuss in detail how 
these arguments depend on the choice of the Lorentz frame. 
To the best of the present author's knowledge, this issue has not been 
previously addressed in the literature.

\section{\label{sec:coh}Non-relativistic neutrinos and coherence conditions}
It is well known that neutrino oscillations can only be observable if neutrino 
production and detection processes cannot discriminate between different 
neutrino mass eigenstates. This is because only flavour eigenstates undergo 
oscillations -- mass eigenstates do not oscillate.%
\footnote{In this paper we shall only deal with neutrino oscillations in 
vacuum. Mass eigenstates can oscillate in matter, but this does not change 
our conclusions about the observability of oscillations of non-relativistic 
neutrinos.}  
Flavour eigenstates are coherent linear superpositions of 
mass eigenstates, and therefore the question of observability of neutrino 
oscillations is closely related to the question of coherence of neutrino 
production, propagation and detection. If at any of these stages  
coherence of different mass eigenstates is violated, oscillations will not 
be observable. We therefore examine here if the coherence conditions are 
satisfied for non-relativistic neutrinos. The answer to this question depends 
on whether the neutrino mass spectrum is hierarchical or quasi-degenerate. 
We first consider the latter case.

\subsection{\label{sec:quasi} Quasi-degenerate in mass non-relativistic 
neutrinos}
Conceptual issues of neutrino oscillation theory can only be consistently 
studied within the quantum-mechanical (QM) wave packet framework 
\cite{nonrel2,beuthe,nuwp1,kayser,Giunti:1991ca,Kiers,Dolgov:1997xr,
Giunti:1997wq,Dolgov:1999sp,Dolgov:2002wy,Giunti1,FarSm,Visinelli:2008ds,
parad,naumov,kerstsmi,nuwp2} or in a formalism based on the quantum field 
theoretic (QFT) approach \cite{Kobzarev:1980nk,Kobzarev:1981ra,Giunti:1993se,
Grimus:1996av,Grimus:1998uh,Ioannisian:1998ch,Cardall:1999ze,beuthe,Beuthe2,
Giunti:2002xg,Dolgov:2004ut,Dolgov:2005nb,AKL1}. 
As has been mentioned above, it 
is usually assumed in these studies that neutrinos are ultra-relativistic. 
However, careful examination of the derivations of the neutrino 
oscillation probability in both QM and QFT-based approaches shows that these 
derivations apply without any modifications also to the case of 
non-relativistic but highly degenerate in mass neutrinos, i.e.\ when 
\be
\frac{|\Delta m^2|}{2E} \ll E\,,
\label{eq:quasi1}
\ee
where $\Delta m^2\equiv m_i^2-m_k^2$ and $E$ are the neutrino mass 
squared difference and the average energy of the neutrino mass eigenstates 
composing a produced neutrino flavour state, respectively. As an example, 
in \cite{parad} it has been demonstrated that the conditions for neutrino 
oscillations to occur and to be described by the standard 
oscillation probability are that ({\em i}\,) the energy difference 
of the neutrino mass eigenstates composing the produced flavour eigenstate, 
\be
\Delta E \equiv \Delta E_{ik}=\sqrt{p_i^2+m_i^2}-\sqrt{p_k^2+m_k^2}\,,
\label{eq:deltaE1}
\ee
is related to their momentum difference $\Delta p\equiv p_i-p_k$, 
mass squared difference $\Delta m^2$ and average group velocity $v_g$ by  
\be
\Delta E \approx 
\frac{\partial E}{\partial p}\Delta p + \frac{\partial{E}}{\partial m^2}
\Delta m^2 = v_g \Delta p + \frac{\Delta m^2}{2E}\,, 
\label{eq:deltaE2} 
\ee
and ({\em ii}\,) the coherence conditions for neutrino production, propagation 
and detection (to be discussed in 
detail in Sections~\ref{sec:prodcoh} 
and \ref{sec:propcoh}) are satisfied. The relation in eq.~(\ref{eq:deltaE2}) 
is satisfied with very good accuracy for ultra-relativistic neutrinos; however, 
it is easy to see that for its validity it is sufficient that 
the neutrino mass eigenstates be quasi-degenerate in mass, i.e.\ condition 
(\ref{eq:quasi1}) be satisfied. In this case different mass eigenstates are 
produced under essentially the same kinematic conditions, i.e.\ their 
energy difference $\Delta E$ and momentum difference $\Delta p$ are small 
compared to their average energy $E$ and average momentum $p$, respectively: 
$|\Delta E| \ll E$, $|\Delta p| \ll p$. This means that one can expand 
$\Delta E$ in $\Delta p$ and $\Delta m^2$, which yields eq.~(\ref{eq:deltaE2}). 

The conditions of coherent neutrino production, propagation and detection 
put upper limits on $\Delta m^2/E^2$ and $\Delta m^2/(2E)$ and 
actually do not require neutrinos to be ultra-relativistic; they 
can also be satisfied if neutrinos are sufficiently degenerate in mass.  

Thus, the standard formalism of neutrino oscillations 
developed for ultra-relativistic neutrinos applies without any modifications  
also to non-relativistic but highly degenerate in mass neutrinos. Such 
neutrinos undergo the usual flavour oscillations provided that the 
standard coherence conditions are satisfied.

In the rest of this paper we shall be assuming that neutrinos are not 
quasi-degenerate in mass. 

\subsection{\label{sec:prodcoh}Non-relativistic neutrinos  and production 
coherence}
Let us now consider neutrino production coherence in the case when 
non-relativistic neutrinos are produced. We start with a general discussion 
of neutrino production coherence. 

\subsubsection{\label{sec:prodcohGen}General case}

Different neutrino mass eigenstates 
can be emitted coherently and compose a flavour state only if the intrinsic QM 
uncertainties of their energies and momenta, $\sigma_E$ and $\sigma_p$, are 
sufficiently large to accommodate their differing energies in momenta.
Assuming absence of certain cancellations  (as will be explained  
towards the end of this subsection), this condition can be written as 
\be
\frac{|\Delta E|}{\sigma_E}\ll 1\,,\qquad\quad
\frac{|\Delta p|}{\sigma_p}\ll 1\,.
\label{eq:deltaEdeltaP}
\ee
If, on the contrary, the uncertainties $\sigma_E$ and $\sigma_p$ are very 
small, then by measuring 4-momenta of the other particles involved in neutrino 
production and using energy-momentum conservation, one could in principle 
determine the energy and momentum of the produced neutrino state with high 
accuracy. 
This would allow one to accurately infer the neutrino mass, which would 
mean that a mass (rather than flavour) eigenstate has been emitted, and 
neutrino oscillations would not take place.   

Indeed, assume that by measuring the energies and momenta of all particles 
taking part in neutrino production we determined 
neutrino energy and momentum with some accuracy. According to QM uncertainty 
relations, the uncertainties of the determined neutrino energy and momentum 
cannot be smaller than the intrinsic energy and momentum uncertainties 
$\sigma_E$ and $\sigma_p$ related to localization of the production process in 
finite space-time region. Assuming $\sigma_E$ and $\sigma_p$ to be independent, 
from the on-shell dispersion relation $E^2=p^2+m^2$ one can then find the 
minimum error in the determination of the squared neutrino mass \cite{kayser}:
\be
\sigma_{m^2}=\big[(2E\sigma_E)^2+(2p\sigma_p)^2\big]^{1/2}.
\label{eq:sigmam2}
\ee  
If $\sigma_{m^2}$ satisfies $\sigma_{m^2}\gg |\Delta m^2|=|m_i^2-m_k^2|$, 
one cannot kinematically distinguish between the mass eigenstates $\nu_i$ and 
$\nu_k$ in the production process, i.e.\ they can be emitted coherently. 
Conversely, for $\sigma_{m^2}\lesssim |\Delta m^2|$ one can 
find out which neutrino eigenstate has actually been produced; this means that 
$\nu_i$ and $\nu_k$ cannot be produced coherently. Since neutrino 
oscillations are a result of interference of amplitudes corresponding to 
different mass eigenstates, absence of their coherence means that no 
oscillations will take place. The flavour transition probabilities would 
then correspond to averaged neutrino oscillations. 

This situation is quite similar to that with the electron
interference in double slit experiments. If 
there is no way to find out which
slit the detected electron has passed through, the detection probability will
exhibit an interference pattern; however if such a determination is possible, 
the interference pattern will be washed out.

In the general case when more than two neutrino species are involved, the mass 
eigenstates $\nu_i$ and $\nu_k$ can be produced coherently if their mass 
difference satisfies $|\Delta m_{ik}^2|\ll \sigma_{m^2}$, even if the other 
pairs of mass eigenstates do not satisfy such a condition and therefore 
cannot be produced coherently. In this case partial decoherence takes place. 

As mentioned above, intrinsic energy and momentum uncertainties 
$\sigma_E$ and $\sigma_p$ characterizing a produced neutrino state are related 
to space-time localization of the production process: the better the 
localization, the larger the $\sigma_E$ and $\sigma_p$, and the easier it is 
to satisfy the production coherence condition. It is instructive to 
formulate this condition in configuration space \cite{kayser,FarSm,parad};  
this will also allow us to find out when eq.~(\ref{eq:deltaEdeltaP}) 
represents the coherent production condition. To this end, consider the 
oscillation phase acquired over the distance $x$ and the time interval $t$ 
from the space-time point at which neutrino was produced:%
\footnote{The phase $\phi_{osc}$ (as well as the energy difference $\Delta E$ 
and the momentum difference $\Delta\vec{p}$) is actually defined for each pair 
of neutrino mass eigenstates $\nu_i$ and $\nu_k$ and should carry the indices 
$ik$. We suppress them to simplify the notation. }
\be
\phi_{osc}=\Delta E \cdot t - \Delta\vec{p}\cdot\vec{x}\,.
\label{eq:oscphase1}
\ee
The 4-coordinate of the neutrino production point has an intrinsic 
uncertainty related to the finite 
space-time extension of the production process; 
this leads to fluctuations of the oscillation phase 
\be
\delta \phi_{osc}=\Delta E \cdot \delta t - \Delta\vec{p}\cdot\delta\vec{x}\,,
\label{eq:oscphase2}
\ee
where $\delta t$ and $|\delta\vec{x}|$ are limited by the duration of 
the neutrino production process $\sigma_t$ and its spatial extension 
$\sigma_X$: $\delta t\lesssim \sigma_t$, $|\delta\vec{x}|\lesssim \sigma_X$. 
For oscillations to be observable, the fluctuations of the oscillation phase 
must satisfy $|\delta\phi_{osc}|\ll 1$ -- otherwise oscillations will be 
washed out upon averaging of the phase over the \mbox{4-coordinate} of 
neutrino production. 
That is, observability of neutrino oscillations requires that the condition 
\be
|\Delta E \cdot \delta t - \Delta\vec{p}\cdot\delta\vec{x}|\ll 1\, 
\label{eq:oscphase2a}
\ee
be satisfied. Barring accidental cancellations between the two terms in 
(\ref{eq:oscphase2a}) and taking into account that 
$\sigma_t\sim \sigma_E^{-1}$, $\sigma_X\sim \sigma_p^{-1}$, we 
arrive at eq.~(\ref{eq:deltaEdeltaP}). Therefore, 

\noindent
{\sl \color{teal}
different neutrino mass eigenstates are produced coherently and hence  
neutrino oscillations may be observable only if the oscillation phase acquired 
over the space-time extension of the production region is much smaller than 
unity.}

\noindent
This condition essentially coincides with the obvious requirement that the 
size of the neutrino production region be much smaller than the oscillation 
length (which corresponds to $\phi_{osc}=2\pi$).  

It should be noted that coherent neutrino production is necessary 
for observability of neutrino oscillations, but it is not by itself 
sufficient: for the oscillations to take place,  
also the propagation and detection coherence conditions must be satisfied. 
Detection coherence can be considered quite similarly to the production one; 
propagation coherence will be discussed in Section~\ref{sec:propcoh}.

\subsubsection{\label{sec:prodcohNonr} Non-relativistic neutrinos}

Let us now discuss the production coherence condition in the case when 
one or more neutrino mass eigenstates are non-relativistic in the frame 
where the neutrino source is at rest or is slowly moving. 
In this case different neutrino mass eigenstates are produced under very 
different kinematic conditions, and therefore have vastly differing energies 
and momenta. We shall demonstrate that large energy and momentum differences 
will then prevent coherent neutrino production. 
 
For illustration, we start with a concrete example. Consider neutrino 
production in 2-body decays at rest  
$X\to l\nu_i$, where $l$ denotes a charged lepton, $\nu_i$ is the $i$th 
neutrino mass eigenstate, and $X$ is either a charged pseudoscalar meson 
($\pi$, $K$, \dots), or $W$-boson, or a charged scalar particle. 
The energies and momenta of the produced neutrino mass eigenstates are
\be
E_i=\frac{m_X^2-m_l^2+m_i^2}{2m_X}\,,\qquad
p_i=\frac{\big\{[m_X^2-(m_l^2+m_i^2)]^2-4m_l^2 m_i^2\big\}^{1/2}}{2m_X}\,.
\label{eq:EiPi}
\ee 
For the energy difference $\Delta E\equiv E_i-E_k$ this gives  
\be
\Delta E= \frac{\Delta m^2}{2m_X}\,.
\label{eq:deltaE3}
\ee
Non-relativistic neutrinos are produced in $X$-boson decay if their mass 
nearly coincides with the energy release $m_X-m_l$. An example is the 
(now defunct) KARMEN time distribution anomaly \cite{karmen}, 
which has been interpreted as a production in $\pi\to\mu\nu$ decay of a 
non-relativistic neutrino with mass $m\simeq 33.9$ MeV and velocity $v\simeq 
0.02$. Assuming that the heaviest of the neutrino mass eigenstates produced in 
$X$-boson decay is non-relativistic and barring near degeneracy of the charged 
lepton and $X$-boson masses, from eq.~(\ref{eq:deltaE3}) we then find  
\be
|\Delta E|\sim m_X\,.
\label{eq:deltaE4}
\ee 

Consider now the energy uncertainty $\sigma_E$ of the produced neutrino 
state. For neutrinos born in decay of a free particle at rest, $\sigma_E$
is given by the energy uncertainty of the parent particle, i.e.\ by its 
decay width $\Gamma_X$. The first of the two coherence conditions in 
eq.~(\ref{eq:deltaEdeltaP}) thus reduces to 
\be
|\Delta E| \ll \Gamma_X\,.
\label{eq:cond1}
\ee
The decay rates $\Gamma(X\to l\nu_i)$ for the processes under discussion are 
given in Appendix A. Their common feature is that they can be written as 
$\Gamma_X=\kappa_X m_X$, where $m_X$ is the mass of the $X$-boson and 
$\kappa_X\ll1$. The smallness of $\kappa_X$ is due to the fact that it 
contains a product of small numerical and dynamical factors; in the cases when 
non-relativistic neutrinos are produced, the coefficients $\kappa_X$ are 
additionally suppressed by a small kinematic factor 
which comes from the suppression of the phase space volume available to the 
final-state particles. We thus have 
\be
\Gamma_X\ll m_X\,. 
\label{eq:cond2}
\ee
{}From eq.~(\ref{eq:deltaE4}) it then follows that the coherent production 
condition (\ref{eq:cond1}) is strongly violated, i.e.\ different neutrino mass 
eigenstates cannot be produced coherently. Note that (\ref{eq:cond2}) is a 
general property of all unstable particles -- their decay width is small 
compared to their mass. The only exception are decays of very broad 
resonances, for which $\Gamma_X\sim m_X$; however, even in this case the 
coherent production condition (\ref{eq:cond1}) is violated if a 
non-relativistic neutrino is involved. 

This result is actually quite general and holds also when neutrinos are 
produced in more complicated decays or in reactions. To see this, recall that 
the produced neutrino state is described by a wave packet, whose 
energy dispersion $\sigma_E$ is determined by the 
temporal localization of the production process. 
The mean energy of the neutrino state $\bar{E}$ can be much larger than 
$\sigma_E$ or of the order of $\sigma_E$, but can never be much smaller than 
the energy dispersion $\sigma_E$. 
For processes with production of a non-relativistic neutrino, the 
differences $\Delta E$ between its energy and 
the energies of the other neutrino mass eigenstates  
are of the order of the corresponding 
mean energies. Therefore, 
\be
|\Delta E| \sim \bar{E} \gtrsim \sigma_E\,,
\label{eq:cond3}
\ee
which means that the first of the two coherent production conditions in 
eq.~(\ref{eq:deltaEdeltaP}) is not met. 

What about the second condition in eq.~(\ref{eq:deltaEdeltaP})? 
When a non-relativistic neutrino is produced, the differences $\Delta p$ 
between its momentum and momenta of the other mass eigenstates satisfy 
$|\Delta p| \gtrsim \bar{p}$, where $\bar{p}$ is the mean momentum, similarly 
to what we found for neutrino energy differences and energy uncertainty. 
However, since momentum is a vector whose projections on coordinate axes can 
be of either sign, one cannot in general claim that the modulus of its mean 
value satisfies $\bar{p}\gtrsim \sigma_p$. In particular, for a wave packet 
describing neutrino at rest, $\bar{p}$ =0 while $\sigma_p$ is finite. 
The momentum dispersion of the produced neutrino state is determined by the 
momentum uncertainty inherent in the production process, which in turn depends 
on the spatial localization of this process. The latter depends on how the 
source particles were created and on other features of neutrino production 
\cite{beuthe}, and there are no simple and general arguments 
that would allow one to tell if the condition $\Delta p\ll \sigma_p$ 
is satisfied for non-relativistic neutrinos, in contrast to the situation 
with the requirement $\Delta E\ll \sigma_E$. However, for slow neutrino 
sources  the temporal duration and spatial localization of the neutrino 
production process are not directly related. This means that in general no 
cancellations between the two terms in (\ref{eq:oscphase2a}) occur, and 
for neutrino production to be coherent both conditions in 
eq.~(\ref{eq:deltaEdeltaP}) must be separately satisfied. Hence, violation of 
the first of these two conditions is sufficient to prevent coherent neutrino 
emission and thus neutrino oscillations. 

One might naturally wonder what happens if we consider the usual neutrino 
oscillations (such as e.g.\ oscillations of reactor, accelerator or atmospheric 
neutrinos)%
\footnote{In what follows by the `usual neutrino oscillations' we shall 
always mean oscillations of neutrinos which are ultra-relativistic in the 
rest frame of their source.}
in a reference frame where one of the neutrino mass eigenstates is 
slowly moving or at rest. Indeed, in that case the relations in 
eq.~(\ref{eq:cond3}) should also be valid, and yet neutrinos must be 
oscillating: the answer to the question of whether neutrinos oscillate 
cannot depend on the choice of the reference frame in which neutrinos are 
considered. We shall discuss this issue in Section~\ref{sec:rest}. 
  
\subsection{\label{sec:propcoh}Wave packet separation and propagation 
coherence}
In addition to neutrino production and detection coherence, there is another 
important coherence condition that has to be satisfied for neutrino 
oscillations to be observable: propagation coherence. 
Coherence may be lost on the way between the neutrino source and detector 
because the wave packets of different neutrino mass eigenstates propagate 
with different group velocities. After long enough time (coherence time) 
they will separate by a distance exceeding the spatial length $\sigma_x$ of 
the wave packets, which then cease to overlap. The coherence time can therefore 
be found from the relation 
\be
|\Delta v_g|\cdot t_{\rm coh}\simeq \sigma_x\,,
\label{eq:cohtime}
\ee
where $\Delta v_g$ is the difference of the group velocities of different 
neutrino mass eigenstates. The corresponding coherence distance is given by 
\be
L_{\rm coh}\simeq v_g\cdot t_{\rm coh}\simeq \frac{v_g}{|\Delta v_g|}\sigma_x\,,
\label{eq:cohlength}
\ee
where $v_g$ is the average group velocity of different neutrino mass 
eigenstates. Since in the case of ultra-relativistic or 
quasi-degenerate in mass neutrinos different mass eigenstates are produced 
under essentially the same kinematic conditions, the lengths of their 
wave packets $\sigma_{xi}$ are practically the same. For processes with 
emission of non-relativistic neutrinos, the lengths of the wave 
packets of different mass eigenstates may differ; the quantity $\sigma_x$ 
in eqs.~(\ref{eq:cohtime}) and (\ref{eq:cohlength}) should then be 
understood as the largest among $\sigma_{xi}$. In all known cases the lengths 
of the neutrino wave packets are tiny (microscopic);%
\footnote{A possible exception is the hypothetical recoilless 
neutrino emission from crystals in M\"{o}ssbauer-type experiments, in which 
$\sigma_{x}$ could actually be as long as a few meters \cite{AKL1}. It is 
not, however, clear if it will ever be possible to realize such experiments.} 
still, in the case of the usual neutrino oscillations (with relativistic 
or highly degenerate in mass neutrinos) the coherence distance $L_{\rm coh}$ 
is macroscopic and very long because $v_g/|\Delta v_g|\simeq 
2E^2/\Delta m^2$ is extremely large. The situation is quite different 
when one or more of the produced neutrinos are non-relativistic in the 
laboratory frame, and neutrinos 
are not quasi-degenerate in mass. The velocity differences between different 
non-relativistic neutrino mass eigenstates (or between relativistic 
and non-relativistic states) in that case are $|\Delta v_g|\sim 1$; the 
coherence 
distance is therefore microscopic, $L_{\rm coh}\sim \sigma_x$. This means that, 
even if non-relativistic neutrinos were produced coherently, they would have 
lost their coherence due to wave packet separation practically immediately, 
before getting a chance of being detected.

\section{\label{sec:lorentz}Lorentz boosts}

We have found that in the case when one or more of the produced neutrino mass 
eigenstates are non-relativistic in the reference frame where their source is 
at rest or is slowly moving, the production coherence condition is violated and 
therefore neutrino oscillations cannot take place. It is interesting to see 
how this analysis changes and what prevents neutrinos from oscillating if we 
go to a frame where all neutrinos are ultra-relativistic. 

A different but related question is this: 
How would the usual neutrino oscillations (such as oscillations of reactor, 
accelerator or atmospheric neutrinos) look like in a reference frame where 
one of the neutrino mass eigenstates is at rest? Neutrinos must obviously 
oscillate in that frame as well, but it is very instructive to see where our 
previous arguments against oscillations of non-relativistic neutrinos fail in 
this case. We study this issue first. 

\subsection{\label{sec:rest}Usual neutrino oscillations in the rest frame 
of $\nu_2$}

Consider for simplicity 2-flavour neutrino oscillations in 1-dimensional 
approach, i.e.\ assuming that $\vec{p}\parallel \vec{x}$. For definiteness, 
we shall assume that neutrinos are produced in pion decays at rest. 
Extensions to the cases 
of more then two flavours and of moving neutrino source are 
straightforward; extension to the full 3-dimensional picture of neutrino 
oscillations is somewhat more involved but does not pose any problems 
\cite{beuthe}. 

Consider first the neutrino oscillation phase in the frame where the 
neutrino source and detector are at rest. 
By making use of eq.~(\ref{eq:deltaE2}) valid for ultra-relativistic neutrinos 
one can rewrite eq.~(\ref{eq:oscphase1}) as   
\be
\phi_{osc}\simeq -\frac{1}{v_g}(x-v_g t)\Delta E +\frac{\Delta m^2}{2p}x\,.
\label{eq:oscphase3}
\ee
The distance $x$ and the time $t$ between neutrino production and detection 
may both be very large, but the difference $x-v_g t$ is always small. It  
vanishes for pointlike neutrinos; in the case when neutrinos are described by 
finite-size wave packets, it is less than or of the order of the spatial 
length of the neutrino wave packet $\sigma_x$: $|x-v_g t|\lesssim \sigma_x$. 
The quantity $\sigma_x$ is, in turn, determined by the space-time extension 
of the neutrino production region and is typically dominated by its temporal 
localization or, equivalently, by the energy uncertainty $\sigma_E$ inherent 
in the neutrino production process \cite{beuthe,Giunti:2002xg,parad}.  
In particular, for neutrinos produced by non-relativistic 
sources $\sigma_x\simeq v_g/\sigma_E$. 
The first term on the right hand side of eq.~(\ref{eq:oscphase3}) is therefore 
$\lesssim \sigma_x|\Delta E|/v_g\simeq |\Delta E|/\sigma_E$. If the first 
of the coherent production conditions in eq.~(\ref{eq:deltaEdeltaP}) is 
satisfied, this term can be neglected, and we obtain the standard expression 
for the oscillation phase $\phi_{osc}\simeq [\Delta m^2/(2p)]x$. 

Let us demonstrate that under very general assumptions the second 
condition in eq.~(\ref{eq:deltaEdeltaP}) actually follows from the first one.
{}From eq.~(\ref{eq:deltaE2}) 
we find that the condition $|\Delta E|/\sigma_E\ll 1$ is equivalent to 
\be
\left|
v_g\frac{\Delta p}{\sigma_E}+\frac{\Delta m^2}{2E\sigma_E}\right|\ll 1\,.
\label{eq:cond4}
\ee
Barring accidental cancellations, this gives 
\be
v_g\frac{|\Delta p|}{\sigma_E}\ll 1\,,\qquad 
\frac{|\Delta m^2|}{2E\sigma_E}\simeq \frac{|\Delta m^2|}{2p}\sigma_x\ll 1\,.
\label{eq:cond5}
\ee
In refs.~\cite{beuthe,Giunti:2002xg,parad} it has been shown that $\sigma_E\le 
\sigma_p$; taking also into account that $v_g\simeq 1$, we find that the 
first strong inequality in (\ref{eq:cond5}) yields the second condition in 
(\ref{eq:deltaEdeltaP}), as advertised. Note that the second condition 
in (\ref{eq:cond5}), $[|\Delta m^2|/(2p)]\sigma_x\ll 1$, has a simple meaning: 
the size of the neutrino wave packet $\sigma_x$ should be much smaller than 
the neutrino oscillation length $l_{osc}=4\pi p/\Delta m^2$. 
In the example we consider (free pion decay at rest), we have $\sigma_E\simeq 
\Gamma_\pi\simeq 2.5\times 10^{-8}$ eV, $E\simeq p\simeq 29.8$ MeV, and for
$\Delta m^2=\Delta m _{\rm atm}^2\simeq 2.5\times 10^{-3}$ eV$^2$ we find that 
the coherent production conditions (\ref{eq:deltaEdeltaP}) are satisfied with 
a very large margin. 
  
Let us now go to a frame where the heavier of the two neutrino mass eigenstates 
(which we choose to be $\nu_2$) is at rest. This is certainly not the best 
frame to consider neutrino oscillations, as the whole setup will look 
rather weird in it! Indeed, assume that in the initial frame where the neutrino 
source and detector are at rest neutrinos are moving in the 
positive direction of the 
$x$-axis. Then in the new frame $\nu_2$ will be at rest, $\nu_1$ will still be 
moving in the positive direction of $x$ (though with a smaller velocity), the 
parent pion will be moving in the negative $x$-direction, and the detector will 
also be moving in the negative direction of $x$ towards the neutrinos. 
On top of that, the wave packet describing the state of $\nu_2$ will be fast 
spreading. Indeed, being in the rest frame of $\nu_2$ means that the mean 
momentum of its wave packet vanishes. Still, the neutrino wave packets are 
characterized by a finite momentum spread, which means that in its rest frame 
$\nu_2$ will have both positive and negative momentum components along 
the $x$-axis, i.e.\ its wave packet will quickly spread. 
Even though this will not affect observability of neutrino oscillations 
because the neutrino detector will ``collide" with neutrinos before a 
significant spreading occurs (see Appendix B),%
\footnote{
The spreading of the wave packets of neutrinos that are ultra-relativistic 
in the rest frame of their source has negligible effect on their oscillations. 
Obviously, the same should also be true in any other frame, including the rest 
frame of one of the neutrino mass eigenstates. 
We discuss these points in Appendix B.}
this adds weirdness to the whole picture. Still, considering neutrino 
oscillations in the rest frame of one of the mass eigenstates is very 
instructive for understanding when and why non-relativistic neutrinos can 
actually oscillate. 

Let us go to a reference frame in which the whole neutrino source -- detector 
setup is boosted with velocity $u$ along the $x$-axis. The standard Lorentz 
transformations read 
\bea
x'=\gamma_u(x+ut)\,,\;~~\qquad t'=\gamma_u(t+ux)\,,~~~
\label{eq:trans1} \\
E_i'=\gamma_u(E_i+up_i)\,,\!\qquad p_i'=\gamma_u(p_i+uE_i)\,,\,
\label{eq:trans2}
\eea
where $\gamma_u=(1-u^2)^{-1/2}$ is the Lorentz factor of the boost and 
the prime refers to the quantities in the new frame. To go to the rest 
frame of $\nu_2$ we choose $u=-v_{g2}=-(p_2/E_2)$, which gives 
$\gamma_u=E_2/m_2$. In the new frame we then have:%
\footnote{Here and below we take into account that neutrinos are 
ultra-relativistic in the original frame, with $E_1\simeq E_2$ and 
$p_1\simeq p_2$.}
\be
E_2'=m_2\,,\qquad\quad p_2'=0\,,\qquad\quad E_1'\simeq 
\frac{m_2^2+m_1^2}{2m_2}\,,\qquad\quad 
p_1'\simeq \frac{m_2^2-m_1^2}{2m_2}\,.
\label{eq:new1}
\ee
For $\Delta E'\equiv E_2'-E_1'$ and $\Delta p'\equiv p_2'-p_1'$ this gives 
\be
\Delta E'\;\simeq\; -\; \Delta p'\;\simeq\; \frac{m_2^2-m_1^2}{2m_2}\,.
\label{eq:new2}
\ee

Next, we consider the transformation laws for neutrino energy and momentum 
uncertainties. For neutrinos produced in pion decay at rest, the energy 
uncertainty is given by the pion decay width: $\sigma_E=\Gamma_\pi$. In a 
moving frame in which the parent  pion has velocity $v_\pi'$, 
the energy uncertainty is given  
by the pion decay width in that frame, 
$\Gamma_\pi'=\Gamma_\pi/\gamma_{v_\pi'}$, where $\gamma_{v_\pi'}=
(1-v_\pi'^2)^{-1/2}$ is the Lorentz factor of the boost from the pion's rest 
frame. That is, upon going from the pion rest frame to a moving frame the 
neutrino energy uncertainty transforms as  
\be
\sigma_E'=\frac{\sigma_E}{\gamma_{v_\pi'}}=\frac{\Gamma_\pi}{\gamma_u}\,,
\label{eq:law1a}
\ee
where we have taken into account that $v_\pi'$ coincides with the boost 
velocity $u$.

Let us consider now the neutrino momentum uncertainty $\sigma_p$. By the 
coordinate -- momentum uncertainty relation, it is the reciprocal of 
the neutrino coordinate uncertainty. 
The latter essentially coincides with the length $\sigma_x$  of the wave 
packet of the produced neutrino. It has been demonstrated in \cite{FarSm,parad} 
that the quantity  $\sigma_{xj}E_j$ is invariant under Lorentz boosts, i.e.\
\be
\sigma_{xj}'=\sigma_{xj}\frac{E_j}{E_j'}=\frac{\sigma_{xj}}
{\gamma_u(1+u v_{gj})}\,,
\label{eq:sigmaxprime}
\ee
where we have used eq.~(\ref{eq:trans2}). For the momentum uncertainty 
$\sigma_{pj}\simeq 1/\sigma_{xj}$ we therefore have
\be
\sigma_{pj}'=\sigma_{pj}
\gamma_u(1+u v_{gj})\,,
\label{eq:sigmapprime}
\ee
i.e. the neutrino momentum uncertainty transforms in the same way as the 
neutrino energy. 

To go from the rest frame of the parent pion to the $\nu_2$ rest frame we 
choose $u=-v_{g2}$, and eqs.~(\ref{eq:sigmaxprime}) and~(\ref{eq:sigmapprime}) 
give 
\be
\sigma_{x2}'=\sigma_{x}\gamma_u\,,\qquad \sigma_{p2}'=
\frac{\sigma_{p}}{\gamma_u}\,.
\label{eq:relat}
\ee 
Here we have taken into account that, although in the pion rest frame (where 
neutrinos are ultra-relativistic) all the 
neutrino mass eigenstates composing the produced flavour eigenstate 
have essentially the same momentum uncertainty $\sigma_p$ and 
their wave packets have the same length $\sigma_{x}$, this is no longer 
true in reference frames where some of the neutrino mass eigenstates are 
non-relativistic. In particular, 
in the rest frame of $\nu_2$ its wave packet is the longest one and therefore 
it is characterized by the smallest momentum uncertainty, 
$\sigma_{p\text{min}}'=\sigma_{p2}'$. Note that is is actually the smallest 
momentum uncertainty that is of interest to us from the viewpoint of possible 
violation of the production coherence condition. 

Combining eqs.~(\ref{eq:new2}), (\ref{eq:law1a}) and (\ref{eq:relat}), 
we find that in the rest frame of $\nu_2$  
\be
\frac{|\Delta E'|}{\sigma_E'}\simeq 
\frac{\Delta m^2}{2m_2}\frac{\gamma_u}{\Gamma_\pi}\simeq \frac{\Delta m^2}
{2E\Gamma_\pi}\gamma_u^2\,,\qquad\quad
\frac{|\Delta p'|}{\sigma_{p\text{min}}'}\simeq 
\frac{\Delta m^2}{2m_2}v_{g2}\frac{\gamma_u}{\Gamma_\pi}\simeq \frac{\Delta m^2}
{2E\Gamma_\pi}\gamma_u^2\,,
\label{eq:new4a}
\ee
where $E\simeq\frac{m_\pi^2-m_l^2}{2m_\pi}$ is the mean neutrino energy in 
the pion rest frame and we have taken into account 
that $\gamma_u=E_2/m_2\simeq E/m_2$. From eq.~(\ref{eq:new4a}) it follows 
that both $|\Delta E'|/\sigma_E'$ and $|\Delta p'|/\sigma_{p\text{min}}'$ 
scale as $\gamma_u^2$. 
Therefore, even though conditions (\ref{eq:deltaEdeltaP}) are satisfied in the 
original frame where the parent pion is at rest, they may be badly violated in 
the rest frame of $\nu_2$ provided that the boost factor $\gamma_u$ is large 
enough, i.e.\ that the group velocity of the second neutrino mass eigenstate 
in the pion rest frame $v_{g2}$ is sufficiently close to 1. 

So, something went wrong here. To understand the root of the problem, let us 
note that the primary condition of coherent neutrino production is the 
requirement (\ref{eq:oscphase2a}) that the variation of the oscillation phase 
with varying 4-coordinate of the neutrino emission point 
be small. Condition (\ref{eq:deltaEdeltaP}) is secondary and obtains from 
eq.~(\ref{eq:oscphase2a}) only under the assumption that the two terms 
in~(\ref{eq:oscphase2a}) are uncorrelated and do not cancel (or approximately 
cancel) each other. 
It is easy to see that it is actually this seemingly innocent assumption that 
led to the above problem.  To show this, let us note that the Lorentz 
transformation (\ref{eq:trans1}) with $u=-v_{g2}\simeq -1$ gives 
\be
\delta t'\simeq \gamma_u(\delta t-\delta x)\,, \qquad\quad
\delta x'\simeq \gamma_u(\delta x-\delta t)\,, 
\label{eq:new7}
\ee
i.e.\ $\delta t'\simeq -\delta x'$. Thus, even if in the original frame 
$\delta t$ and $\delta x$ are completely independent, 
the corresponding quantities in the rest frame of $\nu_2$ are highly 
correlated. In addition, eq.~(\ref{eq:new2}) tells us that 
$\Delta E'\simeq -\Delta p'$.%
\footnote{
While eq.~(\ref{eq:new2}) is specific to neutrinos produced in pion decays 
(or more generally in 2-body decays), the fact that in the new frame 
$\Delta E'\simeq -\Delta p'$ is actually quite general. It directly follows 
from the Lorentz transformation~(\ref{eq:trans2}) with $u\simeq -1$.}
Therefore, in the rest frame of $\nu_2$ 
the two terms in (\ref{eq:oscphase2}) approximately cancel each other: 
\be
\delta\phi_{osc}'=\Delta E'\cdot\delta t'- \Delta p'\cdot\delta x'\simeq 
\Delta E'\cdot(\delta t'+\delta x')\simeq 0\,.
\label{eq:new8}
\ee
This shows that $(i)$ eq.~(\ref{eq:oscphase2a}) does not lead to the 
conditions in eq.~(\ref{eq:deltaEdeltaP}) in this case and ($ii$) no 
enhancement of $\delta\phi_{osc}'$ actually occurs. 
More accurate calculation taking into account the small deviation of 
$u=-v_{g2}$ from  $-1$ yields $\delta\phi_{osc}'=\delta\phi_{osc}\ll 1$, 
so that the coherent neutrino production condition is satisfied in both 
frames. 

This is exactly as it must be: both the oscillation phase and its 
variation, being products of two 4-vectors, are Lorentz invariant. So must 
be the coherence conditions: the answer to the question of whether different 
mass eigenstates are emitted coherently  
cannot depend on the choice of the Lorentz frame in which we look at 
neutrinos. The conditions in eq.~(\ref{eq:deltaEdeltaP}), which are often 
used in the literature as the coherent production conditions, 
are not Lorentz invariant; they follow from the Lorentz 
invariant condition (\ref{eq:oscphase2a}) only  in 
reference frames where the neutrino source is non-relativistic. Obviously, 
they cannot be automatically extrapolated from one Lorentz frame to another. 

So, we can now answer the question posed at the end of 
Section~\ref{sec:prodcohNonr}. In the reference frame in which neutrino 
source is at rest or is slowly moving the two terms in the expression 
$\delta \phi_{osc}=\Delta E \cdot \delta t - \Delta p \cdot\delta x$ 
do not in general cancel, and the coherent production 
condition~(\ref{eq:oscphase2a}) reduces to~(\ref{eq:deltaEdeltaP}). 
Since the usual neutrinos produced in pion decay at rest are highly 
relativistic with very small energy and momentum differences of their 
mass eigenstates, the coherence conditions (\ref{eq:deltaEdeltaP}) are very 
well satisfied for them. In the frame where one of the produced neutrino 
mass eigenstates is at rest, the energy and momentum differences of 
neutrino mass eigenstates become large, and conditions 
(\ref{eq:deltaEdeltaP}) are no longer satisfied, as discussed in 
Section~\ref{sec:prodcohNonr}. However, in this case 
eq.~(\ref{eq:deltaEdeltaP}) does not represents the coherent production 
condition and is actually irrelevant. This happens because the boost with a 
very large Lorentz factor which is necessary to go to the new frame 
leads to near cancellation of the two terms in eq.~(\ref{eq:oscphase2a})  
in that frame. As a result, conditions (\ref{eq:deltaEdeltaP}) no 
longer follow from the coherent production condition~(\ref{eq:oscphase2a}).

\subsection{\label{sec:nonrboost}Boosting non-relativistic neutrinos}

After we have studied in great detail coherence of the usual neutrino 
oscillations in the rest frame of one of the neutrino mass eigenstates, 
it is easy to understand what happens when neutrinos produced as 
non-relativistic in the rest frame of their source are boosted to become 
relativistic. In the original (laboratory) frame, the variations of temporal 
and spatial coordinates of the neutrino production point 
within the production region are not correlated, and 
neither are the energy and momentum differences of the 
neutrino mass eigenstates. Under these circumstances the 
coherent production condition~(\ref{eq:oscphase2a}) leads to   
eq.~(\ref{eq:deltaEdeltaP}). Violation of the first of the constraints 
in (\ref{eq:deltaEdeltaP}), $|\Delta E|\ll \sigma_E$, which, as discussed in 
Section~\ref{sec:prodcohNonr}, takes place in this case, therefore means that 
the coherent production condition ~(\ref{eq:oscphase2a}) is not met. 

Assume now we go to a fast moving frame in which all neutrino mass eigenstates 
are highly relativistic and have nearly the same energies and momenta. 
Because of Lorentz invariance of $\delta\phi_{osc}$, the production 
coherence condition will be violated in the new frame as well. 
Thus, boosting neutrinos that were non-relativistic in the laboratory frame to 
make them relativistic will not let them oscillate, as expected.

\section{\label{sec:disc}Summary and discussion}

We have studied in detail the question of whether neutrinos that are 
non-relativistic in a reference frame in which their source is at rest or 
is slowly moving can oscillate. The answer to this question depends on the 
neutrino mass spectrum. If neutrinos are highly degenerate in mass, the 
standard formalism of neutrino oscillations applies to them without any 
modifications, and they do oscillate provided that the standard coherence 
conditions are satisfied. This also answers the question why non-relativistic 
neutral $K$, $B$ and $D$ mesons oscillate: this is because their corresponding 
mass eigenstates are highly degenerate in mass. 

If, however, non-relativistic neutrinos are not quasi-degenerate 
in mass, their large energy and momentum differences prevent different mass 
eigenstates from being produced coherently. As a result, no oscillations 
with participation of non-relativistic neutrinos are possible. 
The flavour transition probabilities would correspond to averaged-out 
oscillations in that case, and in particular survival probabilities 
would exhibit a constant suppression.%
\footnote{
When the production (or detection) coherence conditions are violated, 
the probability of the overall neutrino production -- propagation -- detection 
process does not factorize into the production rate, oscillation probability 
and detection cross section, so that the very notion of the oscillation 
probability loses its sense. In that case one could still, in principle,  
study the oscillatory behaviour of the overall probability 
(with or without lepton flavour change) as a 
function of the distance between the neutrino production and detection points. 
Decoherence, however, means that no such oscillatory behaviour will take place.}

We have also shown that even if non-relativistic neutrinos were produced 
coherently, they would have lost coherence due to their wave packet separation 
practically immediately, at microscopic distances from their birthplace. 
Although propagation decoherence may in principle be undone by a very 
coherent neutrino detection \cite{Kiers}, 
in the case of non-relativistic neutrinos this would require a completely 
unrealistic degree of coherence of the detection process. 
In addition, even though in general detection may 
restore neutrino coherence if it was lost on the way between the source and 
the detector, the coherence can never be restored if it was violated at 
neutrino production. 

We have also considered in detail how the choice of the Lorentz frame 
influences our arguments and explicitly demonstrated that the coherence 
conditions are Lorentz invariant, as they should be. In particular, since 
neutrinos which are non-relativistic in the rest frame of their source are 
produced incoherently and do not oscillate in that frame, 
they will also be incoherent and will not oscillate upon a boost to a 
reference frame where they are all ultra-relativistic. On the other hand, 
the usual neutrinos that are ultra-relativistic and oscillate 
in the frame where their source is at rest or is slowly moving will maintain 
their coherence and will be oscillating also in the rest frame of any of the 
neutrino mass eigenstates. 

Our discussion demonstrated that the conditions $|\Delta E|/\sigma_E\ll 1$, 
$|\Delta p|/\sigma_p\ll 1$ that are often employed as criteria of 
neutrino production coherence are not Lorentz invariant and should be 
used with caution. They can only serve as the coherent production conditions 
in the case of non-relativistic neutrino sources, and in general should be 
replaced by the Lorentz-invariant constraint on the variation of the 
oscillation phase over the neutrino production region (\ref{eq:oscphase2a}). 

The main reason why neutrinos 
that are non-relativistic in the frame where their source is at rest or 
is slowly moving do not oscillate is their very large energy and 
momentum differences, which significantly exceed the corresponding 
energy and momentum uncertainties inherent in the neutrino production 
process. This is very similar to the reason why charged leptons do not 
oscillate \cite{chlept}.

The results of our study are for the most part in agreement with 
Hinchliffe's rule~\cite{hin}.

\appendix
\renewcommand{\theequation}{\thesection\arabic{equation}}
\appsection
\renewcommand{\thesection}{\Alph{section}}
\section*{Appendix \Alph{section}: Decay rates 
}
We present here the rates of 2-body decays $X\to l\nu_i$, where $l$ denotes a 
charged lepton, $\nu_i$ stands for $i$th neutrino mass eigenstate, and 
$X$ is either a charged pseudoscalar meson ($\pi$, $K$, 
\dots), or $W$-boson, or a charged scalar particle. 
All the rates are given in the rest frame of the parent particle and 
are calculated to leading order in electroweak interaction,  
initially without neglecting any masses of the involved particles. 

We start with the rate of the charged pion decays $\pi\to l\nu_i$.
Direct calculation yields 
\be
\Gamma(\pi\to l\nu_i)=
\frac{g^4}{256\pi}\frac{m_\pi^4}{m_W^4}\frac{f_\pi^2}{m_\pi^2}m_\pi 
|U_{l i}|^2\frac{m_l^2+m_i^2}{m_\pi^2}
\!\left(1-\frac{m_l^2+m_i^2}{m_\pi^2}\right)
\!\!\left\{\left(1-\frac{m_l^2+m_i^2}{m_\pi^2}\right)^2\!-\frac{4 m_l^2 m_i^2}
{m_\pi^4}\right\}^{1/2}\!.
\label{eq:gammapi1}
\ee
Here $U_{li}$ is the element of the leptonic mixing matrix, $g\simeq 0.65$ is 
the $SU(2)_L$ gauge coupling constant, $m_W$ is the $W$-boson mass, 
$f_\pi\simeq 130$ MeV is the pion decay constant, the rest of notation being 
self-explanatory. Note that the pion decay rate is usually expressed through 
the Fermi constant $G_F=\sqrt{2}g^2/(8m_W^2)$; we prefer to express it here 
through the dimensionless gauge coupling constant $g$. 
For other charged pseudoscalar 
bosons ($X=K$, $B$, \dots) the decay rates 
$\Gamma(X\to l\nu_i)$ can be obtained from (\ref{eq:gammapi1}) by the 
obvious substitution $m_\pi\to m_X$, $f_\pi\to f_X$.

Usually, the decay rates of charged pseudoscalar mesons are calculated 
under the assumption that all neutrino masses are very small and can be 
neglected from the outset. The $X\to l \nu_l$ decay rates are then obtained 
by summing over all the neutrino mass eigenstates. The resulting expressions 
are independent of $U_{li}$ due to unitarity of the leptonic mixing matrix. 
Such an approximation is not applicable if neutrinos with mass $m_i\sim m_X$ 
exist. 

For small lepton masses the factor $\frac{m_l^2+m_i^2}{m_\pi^2}$ 
in~(\ref{eq:gammapi1}) describes chiral suppression of the $X$-meson decay; 
however, for decays with production of non-relativistic neutrinos this factor 
is not small, i.e.\ there is no chiral suppression. The factor in the curly 
brackets 
in eq.~(\ref{eq:gammapi1}) (and similar factors in eqs.~(\ref{eq:gammaW1}) 
and (\ref{eq:gammaphi1}) below) is of kinematic origin; it is just the 
magnitude of the momentum of the produced neutrino (and of equal in magnitude 
but opposite in sign momentum of the charged lepton) in units 
of the mass of the parent particle. It vanishes when $m_l+m_i$ approaches the 
parent particle's mass. 

Consider next leptonic decays of $W$ boson. The leading order  
$W\to l \nu_i$  decay rate \vspace*{1.2mm}reads 
\begin{align}
\Gamma(W\to l \nu_i) = \frac{g^2}{48\pi} m_W 
|U_{li}|^2 
\bigg(1-\frac{m_l^2+m_i^2}{2m_W^2}
-\frac{(m_l^2-m_i^2)^2}{2m_W^4} \bigg)
\bigg\{\Big(1-\frac{m_l^2+m_i^2}{m_W^2}\Big)^2-\frac{4 m_l^2 m_i^2}
{m_W^4}\bigg\}^{1/2}. 
\label{eq:gammaW1}
\end{align}

Finally, we consider the rate of decay of a charged scalar $\phi$ caused by 
the Yukawa-type interaction 
\be
{\cal L}_{int} = y\bar{l}\nu_i\phi + h.c.\,,
\label{eq:Lint}
\ee
where $y$ is the Yukawa coupling constant. Note that such charged scalars 
exist in many extensions of the Standard Model, e.g.\ in 2 Higgs doublet 
models. Direct calculation to the leading order in the Yukawa coupling $y$ 
yields 
\be
\Gamma(\phi\to l\nu_i)=\frac{|y|^2}{8\pi} m_\phi\bigg(1-\frac{(m_l+m_i)^2}{m_\phi^2}\bigg)
\bigg\{\Big(1-\frac{m_l^2+m_i^2}{m_\phi^2}\Big)^2-\frac{4 m_l^2 m_i^2}
{m_\phi^4}\bigg\}^{1/2}.
\label{eq:gammaphi1}
\ee

The production coherence condition (\ref{eq:cond1}) is more easily satisfied 
for larger values of $\Gamma_X$; the latter are generally 
increased with decreasing mass of the produced charged lepton $m_l$ (because 
this increases the phase space volume available to the final-state particles). 
It therefore may be useful to consider 
the decay widths of $X$-bosons also in an (unrealistic) limit $m_l\to 0$. 
The rates in eqs.~(\ref{eq:gammapi1}), (\ref{eq:gammaW1}) and  
(\ref{eq:gammaphi1}) then simplify to  
\be
\Gamma(\pi\to l\nu_i)=
\frac{g^4}{256\pi}\frac{m_\pi^4}{m_W^4}\frac{f_\pi^2}{m_\pi^2}m_\pi 
|U_{l i}|^2\frac{m_i^2}{m_\pi^2}
\!\left(1-\frac{m_i^2}{m_\pi^2}\right)^2,\qquad\qquad\;
\vspace*{-4mm}
\label{eq:gammapi1a}
\ee
\begin{align}
\Gamma(W\to l \nu_i) = 
\frac{g^2}{48\pi} m_W 
|U_{li}|^2 
\Big(1-\frac{m_i^2}{m_W^2}\Big)
\bigg(1-\frac{m_i^2}{2m_W^2}
-\frac{m_i^4}{2m_W^4} \bigg),\,
\label{eq:gammaW1a}
\end{align}
\be
\Gamma_\phi=\frac{|y|^2}{8\pi} m_\phi\bigg(1-\frac{m_i^2}{m_\phi^2}\bigg)^2.
\hspace*{6.55cm}
\label{eq:gammaphi1a}
\ee

\appsection
\renewcommand{\thesection}{\Alph{section}}
\section*{Appendix \Alph{section}:
Neutrino wave packet spreading}

Consider the spreading of the neutrino wave packets in the case of the 
usual neutrino oscillations. We shall discuss how the effects of  
this spreading change when going from the 
rest frame of the neutrino source (where all neutrinos are 
ultra-relativistic) to the rest frame of one of the neutrino mass 
eigenstates.

The wave packet spreading is caused by the velocity dispersion, i.e.\ by the 
dependence of the group velocity $\vec{v}_{gi}$ of the neutrino mass 
eigenstate $\nu_i$ on its momentum. Indeed, from $E_i=(\vec{p}^{\,2}+
m_i^2)^{1/2}$ it follows that for $m_i\ne 0$ the group velocity $\vec{v}_{gi}= 
\partial E_i/\partial \vec{p}=\vec{p}/E_i$  is a function of $\vec{p}$. 
Therefore, the momentum spread within the 
neutrino wave packet means that its different momentum components propagate 
with different velocities, leading with time to its spreading. 
The spreading velocity is thus%
\footnote{
We use superscripts to label the Cartesian components of the vectors,  
whereas lower indices are used to mark the 
mass eigenstates. In eq.~(\ref{eq:spread1}) the latter are 
omitted in order not to overload the notation.}
\be
{\rm v}^j\simeq \sum_k\frac{\partial v_g^j}{\partial p^k}\sigma_{p}^k=
\sum_k\frac{1}{E}\big(\delta^{jk}-v_g^j v_g^k\big)\sigma_p^k\,,
\label{eq:spread1}
\ee
where  $\sigma_{p}^k$ is the neutrino momentum dispersion in the $k$th 
direction. For spreading of the wave packet of the $i$th neutrino mass 
eigenstate in the direction of the neutrino propagation (longitudinal 
spreading) we obtain  
\be
{\rm v}_i^\parallel=\frac{m_i^2}{E_i^3}\sigma_{p}\,,
\label{eq:spread2}
\ee
where $\sigma_p$ is the momentum dispersion in the longitudinal direction and 
we have taken into account that for neutrinos that are ultra-relativistic in 
the rest frame of their source 
the momentum dispersion of the different neutrino mass eigenstates 
is practically the same. 

As follows from eq.~(\ref{eq:spread2}), for ultra-relativistic neutrinos the 
longitudinal spreading velocity is very small. As a result, 
the spreading of their wave packets is of no relevance to neutrino 
oscillations. To show this, let us define the characteristic spreading time 
$t_{{\rm spr}\,i}$ as the time over which the wave packet of $\nu_i$ spreads 
to about twice its initial length, $\sigma_x\sim 1/\sigma_p$: 
${\rm v}_i^\parallel t_{{\rm spr}\,i}\simeq 1/\sigma_p$. This gives 
\be
t_{{\rm spr}\,i}\simeq \frac{E_i^3}
{m_i^2 \sigma_p^2}\,.
\label{eq:spread3}
\ee
Let us compare this time with the oscillation time $t_{\rm osc}$ (which 
for ultra-relativistic neutrinos coincides with the oscillation length 
$l_{\rm osc}=4\pi E/\Delta m^2$): 
\be
\frac{t_{{\rm spr}\,i}}{t_{\rm osc}}\simeq \frac{E^2}{4\pi \sigma_p^2}
\frac{\Delta m^2}{m_i^2}\,.
\label{eq:spread4}
\ee
Barring quasi-degeneracy of the neutrino masses, from the oscillation data 
it follows that $\Delta m^2/m_i^2\gtrsim \Delta m_{21}^2/\Delta m_{31}^2 
\sim 1/30$. Next, we note that in realistic situations neutrino energy 
is always very large compared to the energy uncertainty: $E\gg \sigma_E$. 
As an example, for $\pi \to \mu\nu$ decay at rest $E\simeq (m_\pi^2-m_\mu^2)/
(2m_\pi)\simeq 29.8$ MeV and $\sigma_E\simeq \Gamma_\pi= 2.5\times 10^{-8}$ eV, 
so that $E/\sigma_E\simeq 1.2\times 10^{15}$. Since for ultra-relativistic 
neutrinos $\sigma_p\simeq \sigma_E$, 
the ratio in eq.~(\ref{eq:spread4})  
is extremely large.%
\footnote{
A possible exception are supernova neutrinos, for which $E/\sigma_E$ can 
be as small as $\sim 10$. However, in this case the spreading of the neutrino 
wave packets is not relevant to neutrino oscillations either \cite{kerstsmi}.}
Thus, it takes a much longer time for the neutrino wave packet 
to spread by about a factor of two than for neutrino oscillation 
probability to reach its first maximum. This means that the effects of 
wave packet spreading on neutrino oscillations can be safely neglected in all 
realistic situations. 

Note that the requirement $t_{{\rm spr}\,i}\gg l_{\rm osc}$ as a 
condition for neglecting the wave packet spreading effects is actually a 
very conservative one. Indeed, for the usual neutrino oscillations, the 
coherent production condition is satisfied with a large margin, which, in 
particular, means that the initial length of the neutrino wave packets 
satisfies $\sigma_x\ll l_{\rm osc}$ (see Section~\ref{sec:rest}). In these 
circumstances the value of $\sigma_x$ has no effect on neutrino oscillations,%
\footnote{Except for neutrino propagation decoherence, for which the finite 
length of the neutrino wave packet $\sigma_x$ is crucial, 
see Section~\ref{sec:propcoh}. However, as was shown in \cite{kerstsmi}, the 
spreading of the neutrino wave packets does not affect the coherence length, 
which is therefore defined by the initial value of $\sigma_x$.} 
and neither will have its doubling.

One naturally expects that, if the wave packet spreading effect on neutrino 
oscillations is negligible in the rest frame of the neutrino source, 
the same will hold in all other Lorentz frames. 
We shall now demonstrate this explicitly. Let us go to the frame where the 
neutrino source moves with a velocity $u$ in the direction of neutrino 
emission. In the new frame eq.~(\ref{eq:spread2}) yields 
\be
{\rm v}_i^{\parallel}{'}=\frac{m_i^2}{{E_i'}^3}
\sigma_p'\simeq 
{\rm v}_i^{\parallel}\frac{1}{\gamma_u^2(1+u v_{gi})^2}\,,
\label{eq:spread2a}
\ee
where we have used eqs.~(\ref{eq:trans2}) and (\ref{eq:sigmapprime}). 
Note that in the rest frame of the neutrino mass eigenstate $\nu_i$ 
(i.e.\ for $u=-v_{gi}$) the longitudinal spreading velocity of its wave packet 
is a factor $\gamma_u^2$ larger than it is in the rest frame of the parent 
pion. 

Next, let us find the characteristic spreading time $t_{{\rm spr}\,i}'$ in 
the moving frame. Eqs.~(\ref{eq:sigmaxprime}) and (\ref{eq:spread2a}) yield 
\be
t_{{\rm spr}\,i}'\simeq 
\frac{\sigma_x'}{{\rm v}_i^{\parallel}{'}}=t_{{\rm spr}\,i} \gamma_u
(1+u v_{gi})\,.
\label{eq:spread3a}
\ee
Let us now compare the spreading times $t_{{\rm spr}\,i}$ and 
$t_{{\rm spr}\,i}'$ with, respectively, the time intervals between the $\nu_i$ 
production and detection in the original frame and in the new frame, 
$\Delta t_i$ and $\Delta t'_i$.%
\footnote{
Note that if we choose the new frame to be the rest frame of the neutrino mass 
eigenstate $\nu_i$, the quantity $\Delta t'_i$ will actually be 
the interval between the neutrino production time and the time when the 
detector ``collides'' with the resting $\nu_i$.}
Taking into account that $\Delta x_i\simeq v_{gi}\Delta t_i$, from the 
Lorentz transformation law (\ref{eq:trans1}) one finds $\Delta t'_i=
\Delta t_i \gamma_u(1+u v_{gi})$. Together with eq.~(\ref{eq:spread3a}) this 
gives 
\be
\frac{t_{{\rm spr}\,i}'}{\Delta t'_i}=\frac{t_{{\rm spr}\,i}}{\Delta t_i}\,.
\label{eq:spread4a}
\ee
Thus, if the wave packet spreading time is much larger than the neutrino 
flight time in the rest frame of the neutrino source, the same will be 
true in any other Lorentz frame, including the rest frame of one of the 
neutrino mass eigenstates. Therefore, the relative effects of the wave packet 
spreading on neutrino oscillations is frame independent, as expected.

\end{document}